\documentclass[preprint2,epsf]{aastex}

\slugcomment{accepted for publication in ApJ}

\shorttitle{WIEN FIREBALL MODEL}
\shortauthors{IWAMOTO \& TAKAHARA}

\begin{document}

\title{WIEN FIREBALL MODEL OF RELATIVISTIC OUTFLOWS IN ACTIVE GALACTIC NUCLEI}

\author{\scshape S. Iwamoto}
\affil{Yukawa Institute for Theoretical Physics, Kyoto University}
\email{iwamoto@yukawa.kyoto-u.ac.jp}

\and

\author{\scshape F. Takahara}
\affil{Department of Earth and Space Science, Graduate School of Science, Osaka University}
\email{takahara@vega.ess.sci.osaka-u.ac.jp}

\date{Received 2003 June 4; accepted 2003 September 27}

\begin{abstract}

We study steady and spherically symmetric outflows of pure electron-positron pair plasma as a possible acceleration mechanism of relativistic jets up to a bulk Lorentz factor of greater than $10$. We assume that at the inner boundary a ``Wien fireball'' is realized, which is optically thick to Compton scattering but thin to absorption and in a Wien equilibrium state between pairs and photons at a relativistic temperature. As was shown by approximate treatments in our previous paper, the Wien fireball results in a relativistic outflow by thermal expansion, and thus problems with pair annihilation and radiation drag can be avoided. 
In this paper we present numerical solutions obtained with a Monte Carlo simulation of radiative transfer in a relativistic flow. Compton scattering, pair annihilation, and pair creation processes are considered in simulating the photon trajectories, and we evaluate the photon distribution function, pair creation rate, and radiative force. The dynamics of the outflow of pairs are consistently solved with radiative force and pair processes by iteration. 
The numerical results basically confirm our previous finding that the formation of powerful relativistic outflows can be obtained by the Wien fireball. Pair plasma is relativistically accelerated, and the radiative force does not work as a drag force but as an accelerating one because of the relativistic beaming effects. Radiation emitted from the photosphere should be observed as MeV peaked emission at infinity with a luminosity on the order of the kinetic power of jets.

\end{abstract}


\section{INTRODUCTION}

The production and bulk acceleration of relativistic jets in active galactic nuclei (AGNs) is one of the most challenging problems in astrophysics. The basic features to be explained are highly relativistic velocity with a bulk Lorentz factor above 10, a huge kinetic power that is almost comparable to the Eddington luminosity, and collimation into a small opening angle (e.g., Ostrowski et al. 1997; Begelman, Blandford, \& Rees 1984). Although many ideas have been proposed ranging from hydrodynamic and magnetic accelerations to a radiative one, each model has difficulties for the explanation of all aspects of the powerful relativistic jets. Thus, there is no consensus on how jets are produced and accelerated.

In our previous work \citep{Iwamoto02}, we proposed a thermal mechanism of bulk acceleration of powerful relativistic outflows on the assumption of a spherical symmetric and stationary flow. We assumed that at the base of the outflow, electron-positron pair plasma is optically thick to Compton scattering but thin to absorption. High-energy photons and pairs are assumed to be in a Wien equilibrium state. The dynamical features of an optically thick pair plasma are quite different from those of the optically thin one considered in the radiative acceleration model \citep{Phinney82,Inoue97}. We named this initial condition of a pair plasma a ``Wien fireball.'' In the radiative acceleration model, it is well known that radiation drag effect \citep{Phinney82,Sikora96} prevents the pairs from obtaining a bulk Lorentz factor above 5, that the greater part of the pairs are annihilated on the way to bulk acceleration, and that most of the radiative power escapes without scattering. 
Recently, \citet{Beloborodov99} studied the behavior of electron-positron outflows generated by photon-photon collisions above the gamma-ray-emitting accretion disks. The resultant flow was only mildly relativistic because of the radiation drag and Compton cooling due to coexisting soft photons in the cylindrical geometry. In contrast, our Wien fireball model supposes that pairs at relativistic temperatures are generated in a compact region with greater compactness and that the effects of the external soft photons can be neglected in the spherical geometry. 
Thus the difficulties of radiation drag and pair annihilation are successfully avoided owing to the strong beaming of radiation and pair production by accompanying high-energy photons. It was shown that the pair plasma is relativistically accelerated by thermal expansion and that most of the pairs survive to infinity.

However, in our previous paper we made several approximations for a simplified treatment of the radiative transfer. There we treated optically thick and thin (to scattering) regimes separately. We assumed that in the optically thick regime photons and pairs form a single fluid while in the optically thin regime photons are free-streaming. We also assumed that the scattering cross section is given by the Thomson one and that the pair creation rate is negligible in the optically thin regime. In this paper we examine more exactly the features of the relativistic outflow that is initiated by the Wien fireball, by employing the Monte Carlo treatment of radiative transfer. 
In the Monte Carlo treatment, a Klein-Nishina cross section is employed without using the Thomson approximation and the radiation field is faithfully treated both in the optically thick and optically thin regimes. In addition, pair creation and annihilation are also treated by the Monte Carlo method. On the Basis of the Monte Carlo simulation, we validly calculate the angular distribution and energy spectrum of photons, the pair creation rate, and the radiative force. 


In \S~\ref{BASIC FEATURES OF WIEN FIREBALL MODEL} we describe basic features of the Wien fireball model. The difference between the ``fireball'' model in gamma-ray bursts (GRBs) and the Wien fireball model is noted. In \S~\ref{FORMULATION} we present basic equations to solve the outflow of the pair plasma and the Monte Carlo treatment of the radiative transfer. In \S~\ref{RESULTS} we show an example of the numerical results of pair plasma outflow and radiative properties, and in \S~\ref{NUMERICAL RESULTS FOR VARIOUS BOUNDARY VALUES} we present the numerical results for various boundary values. Finally, in \S~\ref{DISCUSSION} we discuss several issues related to the Wien fireball model.

\section{BASIC FEATURES OF WIEN FIREBALL MODEL}
\label{BASIC FEATURES OF WIEN FIREBALL MODEL}

We assume that a pure electron-positron pair plasma, which is optically thick to Compton scattering, is produced at the base of jet. Such a pair plasma will be accompanied by high-energy photons because electron-positron pairs are considered to be mainly produced by photon-photon collisions. Through scattering and pair processes, pairs and photons are assumed to be in a certain equilibrium state called the ``Wien equilibrium'' \citep{Svensson84}. 

\subsection{\it Wien Fireball Model for AGN Jets}

For relativistic bulk acceleration, a huge amount of energy that exceeds the rest-mass energy is required per particle. As such, the fireball model is widely accepted for the bulk acceleration model of cosmic GRBs \citep{Rees92}. The fireball is a compact high-entropy plasma that consists of photons, electron-positron pairs, and a small amount of baryons. It is optically thick both to scattering and to absorption. It is in a complete thermal equilibrium, and the bulk acceleration should be achieved by thermal expansion. In the situation of a fireball, electron-positron pairs, photons, and baryons are dynamically coupled together and there is no need to worry about the radiation drag effect because they expand together as a single fluid. In the fireball model of GRBs, electron-positron pairs are almost wholly annihilated in the course of the thermal expansion. This is because the temperature decreases below the rest-mass energy of the electron before the electron-positron pair creation/annihilation processes become frozen. When the thermal expansion lasts until all the internal energy is expended, the attainable bulk Lorentz factor of the outflow is simply given by $\Gamma_{\rm \infty} \sim {\dot{E} / (m_{\rm p} \dot{N}) }$, which represents the mean free energy per baryon normalized by its rest mass $m_{\rm p}$. Here $\dot{E}$ and $\dot{N}$ denote the total energy flux and total number flux of baryons, respectively, and we use the unit of $c=1$($c$ is the speed of light). For steady, spherically symmetric, and relativistic flows, the bulk Lorentz factor and temperature behave as $\Gamma = \Gamma_0 (r/r_0)$ and $\theta=\theta_0 (r/r_0)^{-1}$, as considered in the fireball model of GRBs \citep{Goodman86,Paczynski86}. Here $r$ is the radius of the fireball, $\theta$ denotes the temperature normalized by electron mass, and the subscript $0$ denotes the physical quantities at the inner boundary. 

\begin{figure}[hbt]
\centering
\epsscale{1.0}
\plotone{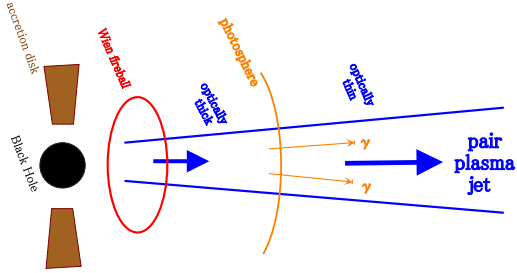}
\caption{
Schematic diagram of a pair plasma flowing out from the optically thick regime to the thin one. In the optically thick regime, pairs and photons behave as a single fluid and expand together. In the optically thin regime, photons stream out almost freely from the photosphere, and pairs and photons do not form a single fluid any longer.
}
\label{model-fig}
\end{figure}

Can this fireball model be directly applied to AGN jets? There is no possibility of a fireball forming because the characteristic size of AGNs is much larger than that of GRBs. The characteristic size of AGNs is $3r_g \sim 10^{14}$ cm, where $r_g \equiv 2GM$ is the gravitational radius for the black hole mass $M$ (the typical value of $M$ is taken to be $10^8$ $M_\odot$), with $G$ being the gravitational constant. If a complete thermal equilibrium is assumed, the temperature of the radiation is estimated as about $10^5$ K, which is far below the rest-mass energy of the electron, and no pair creation is expected. The rest-mass density of baryons in an optically thick plasma is far greater than the energy density of photons, and no relativistic flow is expected. However, if the radiation is optically thin to absorption, a complete thermal equilibrium is not realized and an alternative possibility exists. From the size and luminosity of AGNs, a pair plasma is expected to be optically thin to absorption but thick to scattering. It is possible to form copious electron-positron pairs when high-energy photons are provided sufficiently. Even for mildly relativistic temperatures, copious pairs can be produced for a finite optical thickness to scattering (see the next subsection). In such a state, photons and electron-positron pairs are not in a complete thermal equilibrium, but they are coupled by electron scattering and pair processes. In this state, the photon density is far below that in the complete equilibrium state at a given temperature. The physical state of such a pair plasma may be characterized by a Wien equilibrium state in the simplest case. The Wien equilibrium state is known to be strongly constrained by thermal balance and pair equilibrium owing to an inefficiency of pair annihilation at relativistic temperatures \citep{Lightman82,Svensson84,Kusunose85}. Such a Wien equilibrium state may be established at the base of AGN jets, which we call a ``Wien fireball'' \citep{Iwamoto02}. Although the Wien fireball is not in a complete thermal equilibrium, the dynamical feature of the Wien fireball is the same as that of the normal fireball in GRBs. Pairs and photons are dynamically coupled to each other and relativistically expand together as a single fluid. While the pair plasma is optically thick at the base, it becomes optically thin after the expansion and electron-positron pairs are dynamically decoupled from photons outside the photosphere, where pairs and photons do not behave as a single fluid any longer. As discussed in our previous paper \citep{Iwamoto02} and in subsection~\ref{Survival of Electron-Positron Pairs} 
below, pairs can survive without excessive annihilation in this 
model.  In Figure \ref{model-fig} we illustrate a schematic diagram of the pair plasma outflow.

\subsection{\it Optical Thickness at the Base of the Jet}

In this subsection we argue that the base of the relativistic jet is expected to be optically thick to scattering. Let us consider the optical thickness of the jet in the compact region where its formation, bulk acceleration, and collimation should occur. We can easily estimate the optical thickness under some basic assumptions. The jet is assumed to be composed of almost pure electron-positron pairs with a solid opening angle $\Omega_{\rm j}$, a bulk flow velocity $\beta$, and a Lorentz factor $\Gamma$. The longitudinal optical thickness is estimated by using the Thomson cross section $\sigma_{\rm T}$ as 
\begin{equation}
\tau \sim {n_{\rm e} \sigma_{\rm T} r \over \Gamma} 
= \left( {4 \pi \over \Omega_{\rm j}} \right) 
{1 \over 2 \langle \gamma_{\rm e} \rangle \Gamma^3 \beta} 
\left( {m_{\rm p} \over m_{\rm e}} \right) 
\left( {r_g \over r} \right) \left( {L_{\rm j} \over L_{\rm Edd}} \right)
.
\label{2-01}
\end{equation}
Here $m_{\rm p}$ and $m_{\rm e}$ denote the rest mass of a proton and an electron, respectively, $\langle \gamma_{\rm e} \rangle$ denotes the mean Lorentz factor of random motion of electron-positron pairs, $n_{\rm e}$ denotes the total number density of electrons and positrons, and $r$ denotes the distance from the central black hole. The kinetic luminosity of the jet and the Eddington luminosity are denoted by $L_{\rm j}$ and $L_{\rm Edd}$, respectively. Because the term $m_{\rm p}/m_{\rm e}$ is $1836$ and because the collimation factor $4 \pi / \Omega_{\rm j}$ is much larger than unity, the optical thickness $\tau$ exceeds $1$ when the other terms are not much smaller than $1$, i.e., when $L_{\rm j}$ is comparable to $L_{\rm Edd}$ and when $\langle \gamma_{\rm e} \rangle$ is not excessively high for $r$ of within several times $r_g$. Note that $\Gamma$ is not large at the base of the jet. Therefore, AGN jets should be optically thick to scattering at the scale of the gravitational radius, where jet formation must occur. By setting $\tau=1$ in equation~(\ref{2-01}), the radius of photosphere is also estimated as 
\begin{equation}
r_{\rm ph} \sim \left( {4 \pi \over \Omega_{\rm j}} \right) 
{1 \over 2 \langle \gamma_{\rm e} \rangle \Gamma^3 \beta} 
\left( {m_{\rm p} \over m_{\rm e}} \right) 
\left( {L_{\rm j} \over L_{\rm Edd}} \right) r_g
.
\label{2-02b}
\end{equation}
The characteristic radius of the photosphere $r_{\rm ph}$ for AGNs is about $10$ - $100 r_g$ when $\Gamma$ of around 10 is attained there. This consideration seems to be fully consistent with the Wien fireball model, in which the bulk Lorentz factor increases in proportion to $r$ in the optically thick regime.

\subsection{\it Survival of Electron-Positron Pairs}
\label{Survival of Electron-Positron Pairs}

For the pair plasma outflow, there has been an argument that pairs disappear in the course of the outflow \citep{Celotti93,Blandford95}. If the copious electron-positron pairs are confined in a compact region, catastrophic pair annihilation should occur. Here we briefly discuss why pairs can survive in the Wien fireball model in spite of this. Conventionally, whether pairs annihilate or survive in the course of the expansion may be assessed by comparing the annihilation time scale $t_{\rm ann} \equiv {n_{\rm e^+} / (\dot{n}_{\rm e^+})_{\rm ann}}$ with the dynamical one $t_{\rm dyn} = {r / (\Gamma \beta)}$. Here $n_{\rm e^+}$ and $(\dot{n}_{\rm e^+})_{\rm ann}$ are the number density of positrons and its annihilation rate, respectively. When $t_{\rm ann}$ is shorter than $t_{\rm dyn}$, the pair annihilation is considered to be effective, and this is written as 
\begin{equation}
n_{\rm e^-} \sigma_{\rm T} {R \over \Gamma \beta} 
> {8 \over 3} \left[ 1+ {2 \theta^2 \over \ln (1.12 \theta + 1.3)} \right]
,
\label{2-05}
\end{equation}
which is almost the same as the optically thick condition for Compton scattering~($n_{\rm e^-} \sigma_{\rm T} R / \Gamma >1$) unless $\theta \gg 1$. Here $n_{\rm e^-}$ is the number density of electrons, and we use the annihilation rate given by \citet{Svensson82} in his equations(57) and (59). This coincidence results from the fact that the cross section of pair annihilation and that of Compton scattering are the same order. Therefore, the condition $t_{\rm ann}<t_{\rm dyn}$ is satisfied in the optically thick regime inside the photosphere. In this regime, sufficient pair creation by high-energy photons compensates for pair annihilation as long as the temperature remains relativistic~($\theta \ge 1$). 
Since the time scale of pair creation is comparable to that of pair annihilation, the number of pairs is not determined by the time scale of pair annihilation. Not only photons injected at the inner boundary but also those produced by pair annihilation contribute to the creation of pairs, so that a quasi-pair equilibrium state is realized.

On the other hand, when $t_{\rm ann} > t_{\rm dyn}$, the pair annihilation is ineffective. The number of electron-positron pairs is almost conserved because the pair processes are frozen. There is no need to keep the temperature relativistic in the optically thin regime, so there is no worry about the disappearance of pairs in the Wien fireball model. In our previous paper, we gave the freezeout radius $r_{\rm fr}$, which represents the location at $t_{\rm ann}=t_{\rm dyn}$. It was shown that $r_{\rm fr}$ is almost the same as the photospheric radius $r_{\rm ph}$ and that the survival of pairs is achieved in the case of the ``relativistic freezeout'' \citep{Iwamoto02}.

\section{FORMULATION}
\label{FORMULATION}

We assume that pure electron-positron pairs and high-energy photons are injected at the inner boundary $r=r_0$. The optical thickness of such a region to electron scattering is large, but that to absorption is smaller than unity. We assume for simplicity that pairs and photons are in a Wien equilibrium state at a temperature $\theta_0$ on the boundary. We fix the bulk Lorentz factor at the boundary as $\Gamma_0=\sqrt{3/2}$ and use the total luminosity as a free parameter. The optical thickness $\tau_0$ is calculated from $\dot{E}$ and $\theta_0$. The number densities of pairs and photons in a Wien equilibrium are related by \citep{Svensson84} 
\begin{equation}
{n_{\rm e^+} n_{\rm e^-} \over {n_{\rm \gamma}}^2} 
= \left[ {K_2(1/\theta) \over 2 \theta^2} \right]^2 \equiv A^2 (\theta)
.
\label{4-02}
\end{equation}
Here $K_I$ is the $I$-th order modified Bessel function of the second kind. For pure electron-positron pair plasmas ($n_{\rm e^+} = n_{\rm e^-}$), the number density ratio is given by $n_{\rm e^+}/ n_{\rm \gamma} = A(\theta)$. The total luminosity $\dot{E}$ is written in terms of the number injection rates of positrons $\dot{N}_{\rm e^+, 0}$, electrons $\dot{N}_{\rm e^-, 0}$, and photons $\dot{N}_{\rm \gamma, 0}$ at the inner boundary (which are integrated over the spherical surface $4 \pi {r_0}^2$) as  
\begin{equation}
\dot{E}  
=  \left[ 3\theta_0 \dot{N}_{\rm \gamma, 0} 
+ \langle \gamma_0 \rangle (\dot{N}_{\rm e^+, 0} + \dot{N}_{\rm e^-, 0}) 
\right] m_{\rm e}
.
\label{4-03}
\end{equation}
Here $\langle \gamma_0 \rangle$ is the mean Lorentz factor of random motion of pairs at the boundary. Using equation(\ref{4-02}), we get $\dot{N}_{\rm e^+, 0}$, $\dot{N}_{\rm e^-, 0}$, and $\dot{N}_{\rm \gamma, 0}$ as 
\begin{equation}
\dot{N}_{\rm e^+, 0} = \dot{N}_{\rm e^-, 0} 
= {A(\theta_0) \over 3 \theta_0 
+ 2 \langle \gamma_0 \rangle A(\theta_0) } {\dot{E} \over m_{\rm e}}
,
\end{equation}
\begin{equation}
\dot{N}_{\rm \gamma, 0} = {1 \over 3 \theta_0 
+ 2 \langle \gamma_0 \rangle A(\theta_0) } {\dot{E} \over m_{\rm e}}
.
\label{4-04}
\end{equation}
These number injection rates are also represented by the number densities $n_{\rm e^+, 0}$ and $n_{\rm \gamma, 0}$ at the boundary as 
\begin{equation}
\dot{N}_{\rm e^+, 0} \equiv 4 \pi {r_0}^2 n_{\rm e^+, 0} \Gamma_0 \beta_0
, \
\dot{N}_{\rm \gamma, 0} \equiv 4 \pi {r_0}^2 n_{\rm \gamma, 0} \Gamma_0 \beta_0
.
\label{4-05}
\end{equation}

Under these boundary conditions at $r=r_0=2r_g$, we solve the dynamics of pairs and radiation. For the dynamics of pair plasma, we solve the equations for a relativistic perfect fluid interacting with radiation, which are described in subsection~\ref{Dynamical Equations for Pairs}. Electrons and positrons are assumed to obey the Maxwell-Boltzmann distribution. For the radiative transfer, we use a Monte Carlo simulation in which the effects of the Klein-Nishina cross section and the  anisotropic distribution of photons are considered. The method of the Monte Carlo treatment is described in subsection~\ref{MONTE CARLO TREATMENT OF RADIATIVE TRANSFER}. In this treatment we can assess the radiative force and the pair creation rate by photon-photon collision precisely. The effects of nonuniformity and the relativistic motion of pairs are also dealt with adequately. The radiative force is obtained by evaluating the energy and momentum losses of photons per unit volume of a fluid element, and the pair creation and annihilation rates are given by the loss and gain of the number of photons. In order to solve the dynamics of pair plasma outflow consistently with the radiative transfer, we iteratively solve the dynamics and the radiative transfer. At first, we simulate trajectories of photons by the Monte Carlo technique for a given flow of pair plasma, which is taken from the approximate numerical solutions of \citet{Iwamoto02}. Then we obtain the radiative force and the net creation rate of pairs and solve the dynamics of pair plasma outflow again. We make several iterations until the solution converges and we finally obtain a consistent solution. 

\subsection{\it Dynamical Equations for Pairs}
\label{Dynamical Equations for Pairs}

We treat the electron-positron pairs as a relativistic fluid suffering from interactions with photons. We consider the energy and momentum transfers by Compton scattering and by pair processes. They are represented as ``radiative force'' $F^\mu$. For a steady and spherically symmetric flow, conservation laws of energy, momentum, and number of pairs are given by 
\begin{equation}
{1 \over r^2} {d \over dr} \left[ r^2 (\rho_{\rm e} + P_{\rm e}) 
\Gamma^2 \beta \right] = F^0
,
\label{3-01}
\end{equation}
\begin{equation}
{1 \over r^2} {d \over dr} \left[ r^2 (\rho_{\rm e} + P_{\rm e}) 
\Gamma^2 \beta^2 \right] + {dP_{\rm e} \over dr} = F^1
\ \ ,
\label{3-02}
\end{equation}
\begin{equation}
{1 \over r^2} {d \over dr} \left( r^2 n_{\rm e^+} \Gamma \beta \right) 
= \dot{n}_{\rm e^+}
,
\label{3-03}
\end{equation}
respectively. Here $\rho_{\rm e}$ and $P_{\rm e}$ are the energy density and pressure of electron-positron pairs, which are given by 
\begin{equation}
\rho_{\rm e} = 2 m_{\rm e} n_{\rm e^+} \langle \gamma \rangle
,
\label{3-05}
\end{equation}
\begin{equation}
P_{\rm e} = 2 m_{\rm e} n_{\rm e^+} \theta
,
\label{3-06}
\end{equation}
respectively. The total number density of electrons and positrons is 
$n_{\rm e^+} + n_{\rm e^-} = 2 n_{\rm e^+}$ (since $n_{\rm e^+}=n_{\rm e^-}$ 
for a pure electron-positron pair plasma), and $\dot{n}_{\rm e^+}$ is the net creation rate of pairs, i.e.,  pair creation rate minus annihilation rate. The pair annihilation rate $(\dot{n}_{\rm e^+})_{\rm ann}$ is calculated as \citep{Svensson82} 
\begin{eqnarray}
(\dot{n}_{\rm e^+})_{\rm ann} 
& = & \langle \sigma_{\rm ann} v \rangle n_{\rm e^+} n_{\rm e^-}  \nonumber \\
& = & {3 \over 8} \sigma_{\rm T} n_{\rm e^+} n_{\rm e^-}  
\left[ 1+ {2 \theta^2 \over \ln(1.12 \theta + 1.3)} \right]^{-1}
,
\label{3-04}
\end{eqnarray}
where $v$ denotes the thermal velocity of pairs. The right-hand side of the energy and momentum conservation laws denotes the radiative force in the coordinate frame, which is given through the Lorentz transformation of the radiative force in the comoving frame of a fluid element. The pair creation rate and radiative force are calculated by the Monte Carlo simulation described in the next subsection. 

Although we solve the optically thick and thin regimes with a 
single method, the notion of the photospheric radius $r_{\rm ph}$, where the optical thickness to the scattering becomes unity, may be important to understand the numerical results. 
It should be noted that $r_{\rm ph}$ depends on the energy and direction of the photon and that most photons are moving outward in the radial direction in the relativistic outflow. Although the energy dependence of the cross section is significant for photons around MeV range, we define the photospheric radius $r_{\rm ph}$ conventionally by using the Thomson cross section as 
\begin{equation}
\tau 
\equiv \int^{\infty}_{r_{\rm ph}} dr \ (n_{\rm e^+} + n_{\rm e^-}) 
\sigma_{\rm T} \Gamma (1-\beta) 
= 1
.
\label{5-01}
\end{equation}
Because of the Klein-Nishina effect, the real photosphere is located at a somewhat smaller distance than this conventional one.

\subsection{\it Monte Carlo Treatment of Radiative Transfer}
\label{MONTE CARLO TREATMENT OF RADIATIVE TRANSFER}

\begin{figure}[h]
\centering
\epsscale{1}
\plotone{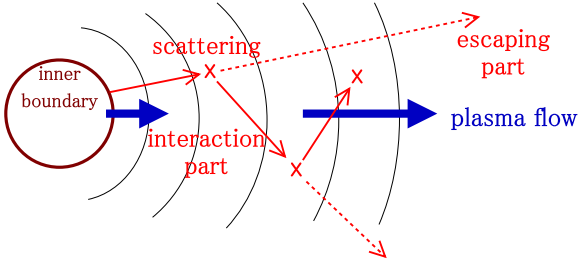}
\caption{The schematic picture of the trajectory of a simulated photon propagating in the pair plasma outflow. Since the flow and photon distribution are highly nonuniform, the spatial region divides into many spherical shell elements.}
\label{mesh}
\end{figure}

The Monte Carlo method for Compton scattering was described in \citet{Pozdnyakov77,Pozdnyakov83}, where the Klein-Nishina cross section is considered. Here we should also take into account pair annihilation and creation, and we need to calculate the distribution of photons, too. In addition, we should take the bulk flow into account. Because the flow velocity is relativistic and nonuniform, we divide the calculation regime of pair plasma flow into many spherical shell elements (see Figure \ref{mesh}), in practice 3323 shells from $r=2r_g$ to $10^4r_g$ with an equal logarithmic interval. We set the physical quantities of the pair plasma (the bulk velocity $\Gamma\beta$, temperature $\theta$, and number density $n_{\rm e^+}$ [$= n_{\rm e^-}$]) to each fluid element. 

We simulate the trajectory of each photon propagating through the pair plasma and photon field, where the photon interacts with electrons, positrons, and photons themselves. Elementary processes to be considered are Compton scattering and electron-positron pair annihilation and creation. While the pair annihilation process generates photons, the pair creation process makes photons vanish, and Compton scattering exchanges photons. We simulate all these processes by the Monte Carlo method. In the simulation, we employ a photon splitting method by assigning a weight to a simulated photon and tracing many trajectories for each injected photon. 

Photons are injected both from the inner boundary and from pair annihilation throughout the flow. At the inner boundary, we generate photons obeying an isotropic Wien distribution in the comoving frame ($\Gamma_0=\sqrt{3/2}$) with the same temperature as the electron-positron pairs. Then, we transform them to the coordinate frame. For the photons arising from pair annihilation, we generate two photons per event. Pair annihilation photons are generated according to the rate given in equation~(\ref{3-04}). We randomly choose the annihilating pairs (electrons and positrons) from the Maxwell-Boltzmann distribution by taking into account the dependence of annihilation probability on the cross section and relative velocity between the annihilating particles. We calculate the energy and momentum of photons from those of annihilated pairs by using conservation laws of energy and momentum. The spectrum of the annihilation photons and the method of calculation are taken from \citet{Ramaty81}.

We generate $\mathfrak{N}_{\rm b}$ simulated photons at the inner boundary; for most cases we take $\mathfrak{N}_{\rm b}=2\times 10^5$. Conforming to the number injection rate of real photons at the inner boundary $\dot{N}_{\rm \gamma, 0}$, the weight parameter of the simulated photons $w_{\rm b}$ is assigned as 
\begin{equation}
w_{\rm b} = {1 \over \mathfrak{N}_{\rm b}} \dot{N}_{\rm \gamma, 0}
.
\label{4-06}
\end{equation}
Each photon is given an equal weight at injection, for simplicity. 

For the pair annihilation photons we generate and inject $2 \mathfrak{N}_{\rm a}=200$ photons per each fluid element (the factor $2$ arises from the fact that each annihilation event generates two photons) so that the total number of simulated annihilation photons is $6.646\times 10^5$ for 3323 fluid elements. We set the weight parameter of the annihilation photons $w_{\rm a}$ as 
\begin{equation}
w_{\rm a} 
= {1 \over \mathfrak{N}_{\rm a}} 
\left[ (\dot{n}_{\rm e^+})_{\rm ann} \ \delta V' {\delta t' \over \delta t} 
\right] 
= {1 \over \mathfrak{N}_{\rm a}} (\dot{n}_{\rm e^+})_{\rm ann} \ \delta V 
.
\label{4-07}
\end{equation}
Here $\delta V'$ and $\delta V$ are the comoving volume and the coordinate volume of the fluid element, respectively. In equation~(\ref{4-07}), the factor $\delta t' / \delta t$ converts the annihilation rate in the comoving frame to that in the coordinate frame. [We note that $(\dot{n}_{\rm e^+})_{\rm ann}$ means the pair annihilation rate not in the coordinate frame but in the comoving frame.] At the second equality, we use the Lorentz invariance $\delta t' \delta V' = \delta t \delta V$. These weight parameters are used when the physical quantities of the radiation field are calculated. The weight parameter is suitably divided into those of split-simulated photons so as to conserve their sum. The weight parameter decreases after the photon-photon absorption. 

At first, we calculate the probability of interaction within the shell $P_{\rm int}$ and that of escaping into the adjacent shell $P_{\rm esc}$. These are given by 
\begin{equation}
P_{\rm esc} = \exp(-\tau_{\rm tot})
, \
P_{\rm int} = 1- P_{\rm esc}
,
\label{8-02-03}
\end{equation}
where $\tau_{\rm tot}$ is the total optical thickness to all the interaction processes (Compton scattering and pair creation; $\tau_{\rm tot}= \tau_{\rm comp}+ \tau_{\rm cre}$) along the trajectory in the shell. Here $\tau_{\rm comp}$ and $\tau_{\rm cre}$ represent the optical thickness for Compton scattering and that for electron-positron pair creation to the shell boundary, respectively. The simulated photon is divided into two parts: the ``escaping'' part and the ``interaction'' part (see Figure~\ref{mesh}). The photon weight of each part is divided in proportion to the probability $P_{\rm esc}$ and $P_{\rm int}$. The trajectory of the escaping part is regarded as entering the adjacent shell and that of the interaction part suffers from further interaction before escape and continues to be simulated to trace the trajectories of photon again and again. 

For Compton scattering, we use the Klein-Nishina cross section $\sigma_{\rm KN}$, and $\tau_{\rm comp}$ is calculated considering the Maxwellian distribution function $f_{\rm e}$ of scattering particles \citep{Pozdnyakov83} as 
\begin{equation}
\tau_{\rm comp} = \Delta l \int d^3p f_{\rm e}\sigma_{\rm KN}(1-v\mu)
,
\end{equation}
where $v$ is the velocity of the particle, $\mu$ is the cosine between the directions of the photon and particle, and $\Delta l$ denotes the path length of the photon trajectory to the boundary of the adjacent shell. 
For pair creation, the cross section for photon-photon collision $\sigma_{\rm \gamma \gamma}$ is taken from the representation in \citet{Gould67}. Optical thickness for pair creation $\tau_{\rm cre}$ is calculated from the distribution function of photons $f_{\rm \gamma}$ as 
\begin{equation}
\tau_{\rm cre} 
= \Delta l \int d\Omega_{\hat{\varepsilon}} \int d \hat{\varepsilon} 
\ f_{\rm \gamma}(\hat{\varepsilon}, \Omega_{\hat{\varepsilon}}) \ 
\sigma_{\rm \gamma \gamma}(\varepsilon, \hat{\varepsilon}, \Theta)v_{\rm R}
.
\end{equation}
Here $\hat{\varepsilon}$ and $\varepsilon$ are the energies of target and incident photons, $\Omega_{\hat{\varepsilon}}$ is the propagation direction of the target photon, $\Theta$ is the collision angle between the target and incident photons, and $v_{\rm R}(\mu) = 1 - \cos\Theta$ is the relative velocity between photons. 

Thus, we get the probabilities for Compton scattering and pair creation (which means photon vanishing): 
\begin{equation}
P_{\rm comp} 
= {\tau_{\rm comp} \over \tau_{\rm comp}+ \tau_{\rm cre}} P_{\rm int}
, \
P_{\rm cre} 
= {\tau_{\rm cre} \over \tau_{\rm comp}+ \tau_{\rm cre}} P_{\rm int}
.
\end{equation}
We choose Compton scattering or pair creation by this ratio of probabilities by random number generation. When we choose pair creation, we make the photon vanish and quit tracking the trajectory of the photon for this part. When we choose Compton scattering, we simulate the Compton scattering and further trace the scattered photon as in \citet{Pozdnyakov83}. Then we continue tracking the trajectory to the next interaction and escape. 

At each interaction point, the weight parameter of the interaction part decreases by a factor of $P_{\rm int}$. If the weight parameter of the interaction part becomes $\epsilon_w$ times smaller than that at the injected point, we finish tracing the photon trajectory. For the nondimensional technical parameter $\epsilon_w$ ($\epsilon_w < 1$), we set at $10^{-12}$. By using the weight parameter like this, photons that have far smaller probability than $1/\mathfrak{N}_{\rm a}$ or $1/\mathfrak{N}_{\rm b}$ can be adequately treated \citep{Pozdnyakov77}.

After simulating $N$ photons (typically we take $N=2\times 10^5$), photon distribution in the phase space is calculated as a superposition of simulated photons. The energy density and flux of photons are obtained. We can also obtain radiative force and pair creation rate by accumulating the energy, momentum, and number losses of simulated photons. While the pair annihilation rate is derived from the distribution of electron-positron pairs, the pair creation rate is derived from that of photons as follows. 
\begin{equation}
(\dot{n}_{\rm e^+})_{\rm cre} 
= {1 \over 2} \int {d \Omega_\varepsilon } \int {d\Omega_{\hat{\varepsilon}}}
 \int d \varepsilon \ f_{\rm \gamma}(\varepsilon) \ 
 \int d \hat{\varepsilon} \ f_{\rm \gamma}(\hat{\varepsilon}) 
 \sigma_{\rm \gamma \gamma} (\varepsilon, \hat{\varepsilon}, \Theta) 
 v_{\rm R}
,
\label{4-01}
\end{equation}
where $\Omega_\varepsilon$ is the direction of the incident photon.

\section{AN EXAMPLE OF NUMERICAL RESULTS}
\label{RESULTS}

We have performed numerical simulations for a wide range of parameters  of the total luminosity $\dot{E}$ and the boundary temperature $\theta_0$. In this section we present a typical example of the numerical results for inner boundary conditions 
\begin{equation}
\dot{E} / L_{\rm Edd} = 24.9 \ \ , \ \ \theta_0 = 2 \ \ , 
\ \ \mbox{with} \ \ 
\tau_0 =85.4
.
\label{5-02}
\end{equation}
We first present the results of the pair plasma outflow and next those of the radiation field. The number, energy, and momentum transfers from the radiation field to the pair plasma are also presented. 

\subsection{\it Pair Plasma Outflow}

\begin{figure}[p]
\centering
\epsscale{0.65}
\plotone{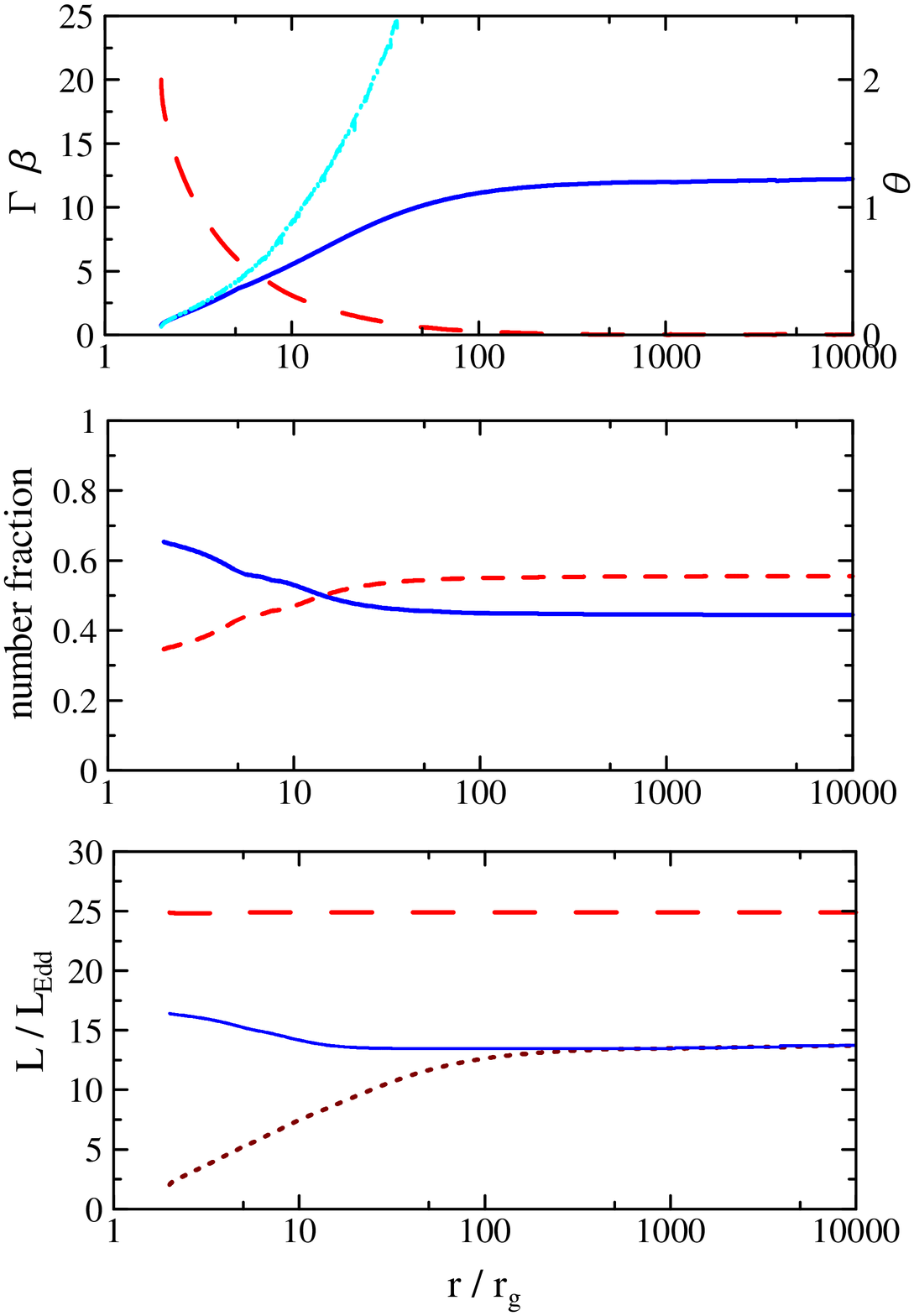}
\caption{
{\it Top:} Behavior of velocity and temperature for $\dot{E}/L_{\rm Edd}=24.9$, $\theta_0=2$, with $\tau_0 = 85.4$. The solid line denotes the bulk Lorentz factor of the pair plasma $\Gamma\beta$, and the dash-dotted line denotes the equilibrium Lorentz factor of radiation field $\Gamma_{\rm eq} \beta_{\rm eq}$. The dashed line denotes the temperature of the pair plasma. {\it Middle:} Fraction of the particle and photon numbers. The solid and dashed lines denote the number fractions of pairs and photons, respectively. {\it Bottom:} Luminosities of various components. The dashed line denotes the total luminosity of pairs and radiation. The solid line denotes the kinetic luminosity of pairs, while the dotted line denotes the luminosity carried in a form of the rest mass of pairs. Thus, the interval between the solid and dotted lines denotes the luminosity carried in the form of the thermal energy of pairs, and that between the dashed and solid lines denotes the luminosity carried by radiation. 
}
\label{velo-frac-lumi2}
\end{figure}

Here we show the result of the pair plasma outflow. The top panel of Figure \ref{velo-frac-lumi2} shows the behavior of the bulk velocity $\Gamma\beta$ and the temperature $\theta$. It is seen that the photosphere is located at $r_{\rm ph}= 9.98 r_g$ and that the temperature rapidly decreases with radius, becoming $\theta_{\rm ph}= 0.311 $ at the photosphere. The bulk Lorentz factor at the photosphere turns out to be $\Gamma_{\rm ph}\beta_{\rm ph}= 5.52$. These numerical values agree well with the analytic prediction of $r_{\rm ph} \simeq 8.84 r_g$, $\theta_{\rm ph} \simeq 0.452$, and $\Gamma_{\rm ph} \simeq 5.41$ \citep{Iwamoto02}. The behavior inside the photosphere represents the basic feature of relativistic thermal expansion, as expected. Outside the photosphere, the pair plasma is further accelerated by the thermal energy at the photosphere and by the additional acceleration by radiative force. The acceleration of bulk motion is saturated at about $r \simeq 100 r_g$ and the terminal Lorentz factor is  $\Gamma = 12.2$. Compared with numerical results in our previous paper, the behavior of the outflow is quite similar except for small differences in numerical values. The location of the photosphere slightly shifts outward. This leads to a slightly smaller $\theta_{\rm ph}$ and a slightly larger $\Gamma_{\rm ph}$. There are also some differences in the asymptotic quantities. Compared to the analytic estimates, the asymptotic bulk Lorentz factor $\Gamma_{\infty}$ is smaller by about $25$\%, the number fraction of electron-positron pairs is larger by about $10$\%, and the kinetic luminosity of pairs is smaller by about $10$\%.

The middle panel of Figure \ref{velo-frac-lumi2} shows the relative number fraction of pairs and photons as a function of the radial coordinate. The pair fraction decreases steadily according to the decrease of temperature. Because pair creation and annihilation processes are unfrozen inside the photosphere, the decrease of temperature leads to the decrease of pair fraction. Outside the photosphere, the decrease of pair fraction is modest and becomes negligible for $r>50r_g$. About $30$\% of the initial pairs annihilate and about $70$\% survive in this calculation. 
Therefore, this result confirms our prediction that the pair annihilation problem can be avoided successfully. The bottom panel of Figure \ref{velo-frac-lumi2} shows the luminosities of pairs and radiation as a function of the radial coordinate. The total luminosity turns out to remain constant by necessity. Inside the photosphere, the kinetic luminosity of pairs slightly decreases because pairs are converted to photons. Outside the photosphere, the kinetic luminosity of pairs is kept almost constant, where the thermal energy of pairs at the photosphere converts to bulk kinetic energy by thermal expansion. Although pairs lose their kinetic luminosity by their annihilation process to some degree, acceleration by radiative force compensates for the annihilation loss of the kinetic luminosity. The terminal kinetic luminosity of pairs turns out to be $52$\% of the total luminosity, while the remaining $48$\% is carried away by photons. The spectrum and the angular distribution of these photons are shown in the next subsection.


\subsection{\it Radiation Field}


\subsubsection{\it Spectrum of Photons}

\begin{figure}[p]
\centering
\epsscale{0.8}
\plotone{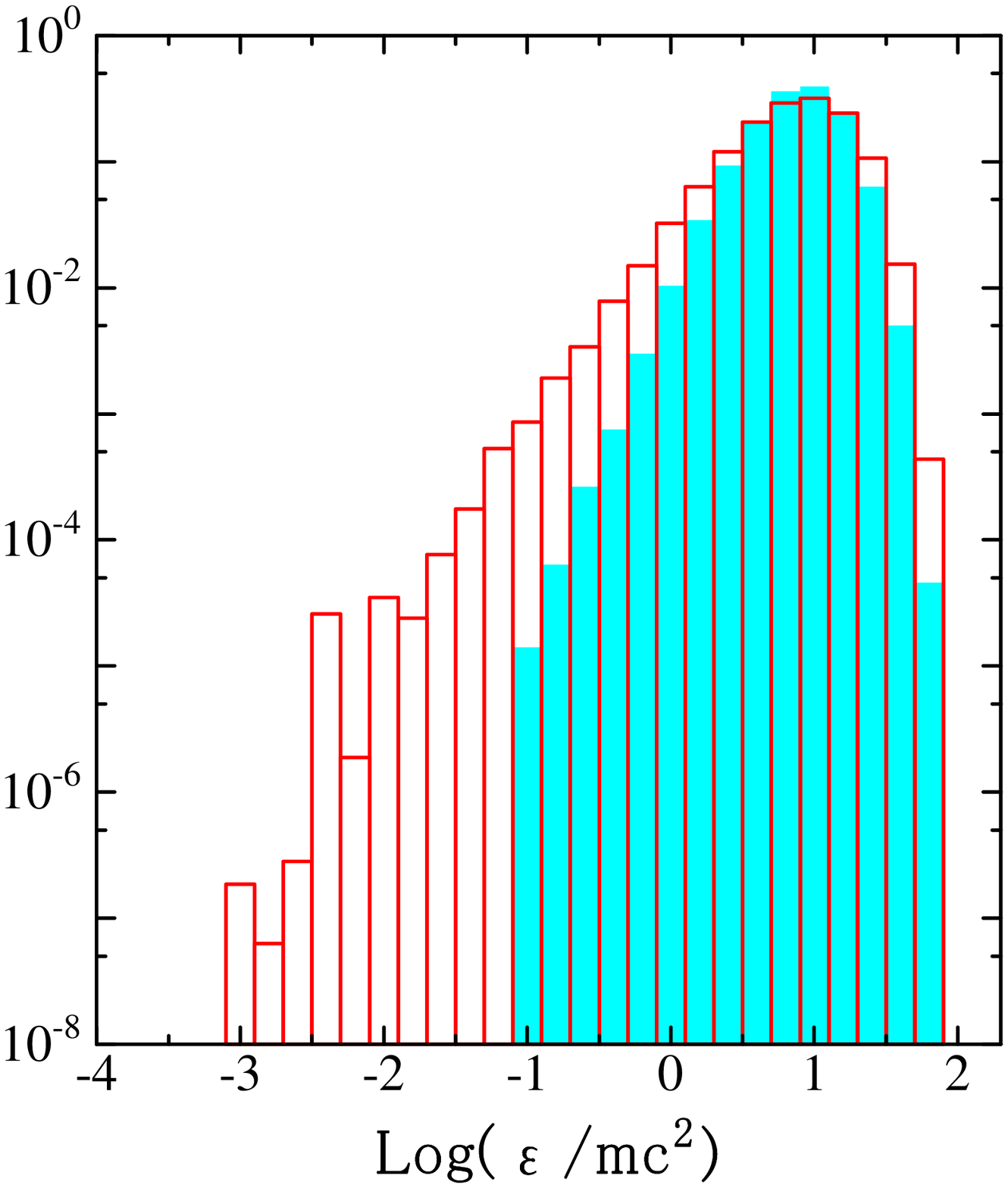}
\caption{
Spectrum of simulated photons. The shaded histogram denotes the spectrum of injected photons at inner boundary $r= r_0 = 2 r_g$, which we call the ``input spectrum.'' The open histogram denotes that of escaping photons from the outer boundary $r = 10^4 r_g$, which we call the ``output spectrum.'' Both spectra are measured in the coordinate~(observer's) frame. Here, $\varepsilon / m c^2$ is the photon energy normalized by the electron rest-mass energy.
}
\label{spectrum}
\end{figure}

In Figure \ref{spectrum} we present the spectrum of photons at the inner boundary ($r_0 = 2r_g$) and at infinity ($r_\infty = 10^4 r_g$), both measured in the coordinate~(observer's) frame. The former is the ``input spectrum,'' which represents the spectrum of injected photons at the inner boundary. The latter is the ``output spectrum,'' which represents the spectrum of escaping photons at the outer boundary. Compton scattering and pair annihilation and creation change the input spectrum into the output one. Basically, the output spectrum is not so different from the injected one, and the injected Wien spectrum of photons is almost conserved. This fact can be comprehended by the energy conservation law per photon. Photons (contained in the Wien fireball) are at a relativistic temperature at the inner boundary, but the bulk motion of the photons is only mildly relativistic. As the Wien fireball expands, the temperature of photons decreases, but at the same time the Lorentz factor of the bulk motion increases, which boosts the energy of photons by the relativistic beaming effect. The effect of decreasing temperature, i.e., the mean energy of photons in the comoving frame, is compensated for by the relativistic beaming effect. In the course of the expansion, the internal energy of the Wien fireball is converted to the kinetic energy of the bulk motion according to the energy conservation law. Thus, the mean energy of photons in the coordinate frame is almost conserved. 


Two features of the output spectrum are noted. First, the number flux of output photons is somewhat larger than that of the input ones because net pair annihilation increases the photon number. Second, the output spectrum is somewhat broader than the input Wien spectrum. The latter is mainly due to Compton scattering in the optically thin regime, which tends to make the pairs and photons depart from an equilibrium. Not only photons injected at the inner boundary but also those injected by pair annihilation are scattered by electrons and positrons. Down-scattering by cooled pairs reduces the energy of photons and forms the low-energy tail of the output spectrum. On the other hand, the high-energy part of output spectrum also enlarges a little, which is also due to Compton scattering. The temperature of electron-positron pairs is not uniform, and scattering by the pairs at multiple temperatures broadens the photon spectrum. The energy dependence of the cross section (Klein-Nishina regime of the cross section) affects the broadening too (at higher energies we see a deeper and higher temperature region). The output spectrum is peaked around a few MeV, and its luminosity is comparable to the kinetic luminosity of pairs.

\subsubsection{\it Angular Distribution of Photons}

\begin{figure}[htp]
\centering
\epsscale{0.7}
\plotone{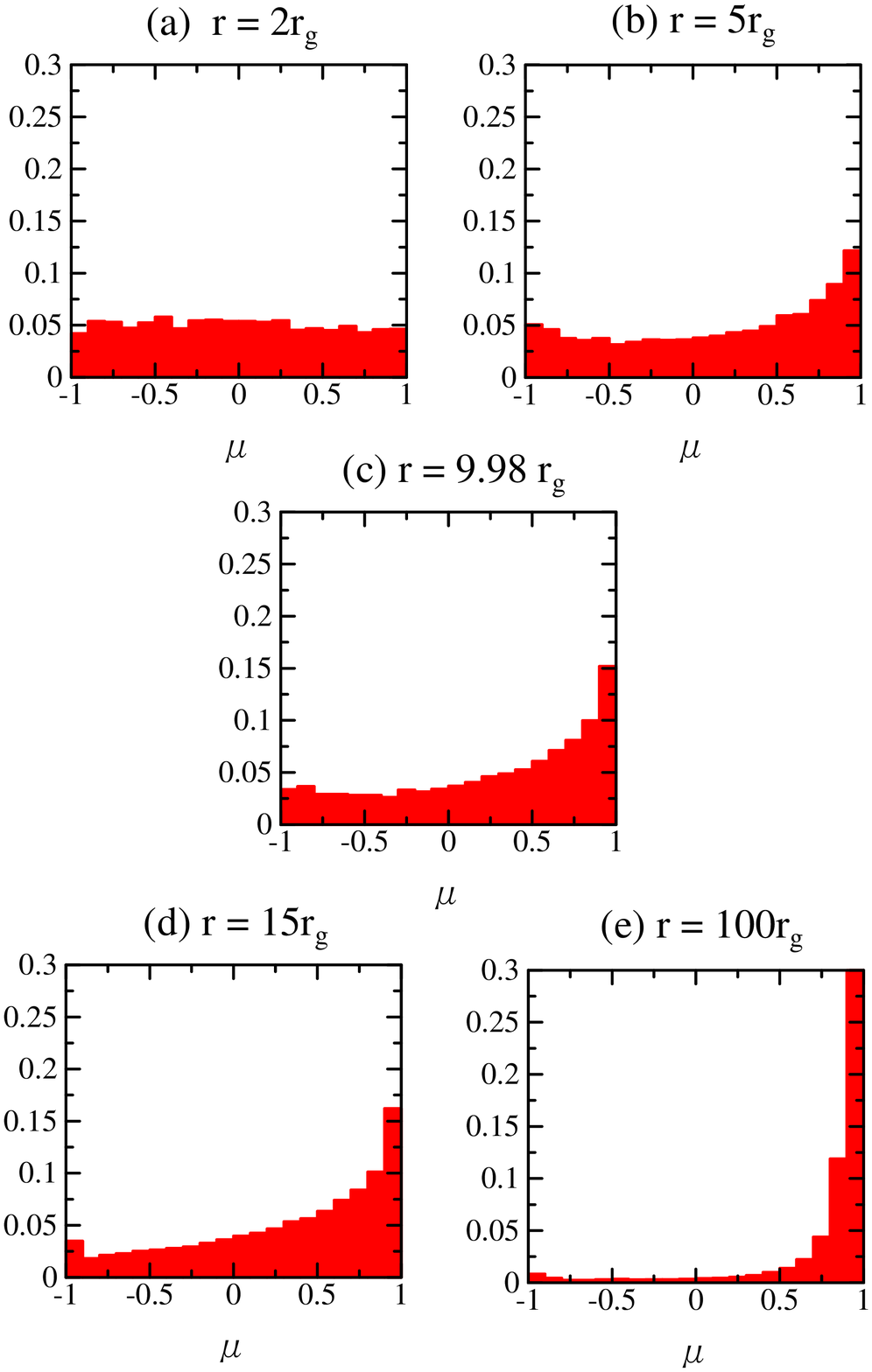}
\caption{
Angular distribution of photons at various shells (measured in the comoving frame of the pair plasma). These histograms represent the fraction of the photon number oriented to $\cos^{-1} \mu$ from the radial direction. The radial coordinates are (a) $r=2r_g$, (b) $r=5r_g$, (c) $r=9.98r_g$, (d) $r=15 r_g$, and (e) $r=100r_g$: (a) and (b) are in the optically thick regime, (d) and (e) are in the optically thin regime, and (c) corresponds to the photosphere. 
}
\label{angular distribution}
\end{figure}

In Figure~\ref{angular distribution} we present the angular distribution of photons in the comoving frame of the pair plasma. Angular distribution is almost isotropic in the optically thick regime. The histograms of photons (Figs. \ref{angular distribution}{\it a} and \ref{angular distribution}{\it b}) show an almost flat distribution. On the other hand, those in the optically thin regime show a skew distribution. As is seen in Figure \ref{angular distribution}{\it d} and \ref{angular distribution}{\it e}, outgoing photons ($\mu>0$) dominate. Obviously, the degree of skewness increases with $r$ (Fig. \ref{angular distribution}{\it a} $\rightarrow$ \ref{angular distribution}{\it b} $\rightarrow$ \ref{angular distribution}{\it c} $\rightarrow$ \ref{angular distribution}{\it d} $\rightarrow$ \ref{angular distribution}{\it e}). In the optically thin regime (Figs. \ref{angular distribution}{\it d} and \ref{angular distribution}{\it e}), there is a small fraction of ingoing photons that are scattered back or emitted inward in the optically thin part. At the photosphere (Fig. \ref{angular distribution}{\it c}), the angular distribution of photons is mildly anisotropic.

\subsubsection{\it Pair Creation and Annihilation Rates}

\begin{figure}[htp]
\centering
\epsscale{0.65}
\plotone{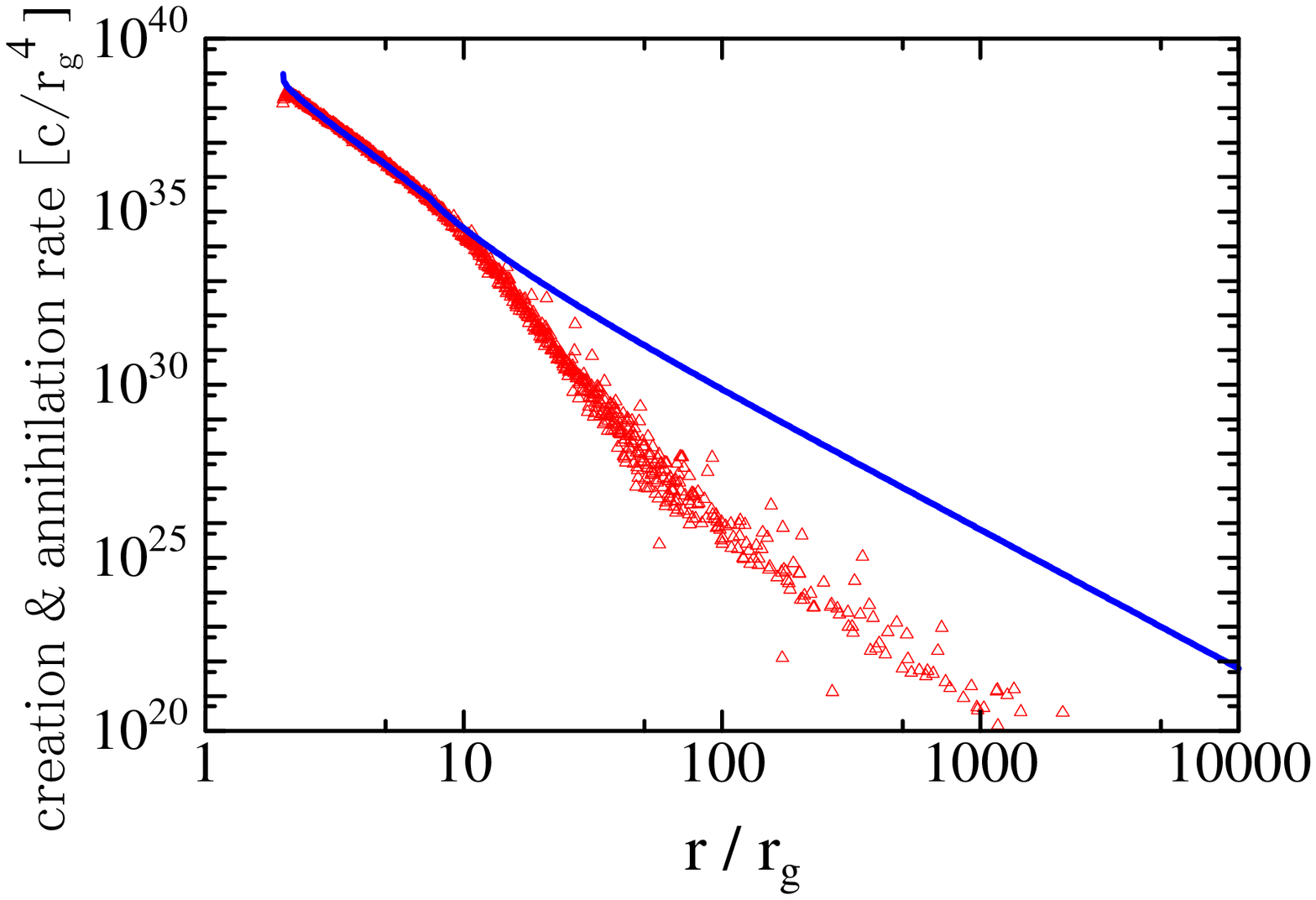}
\caption{
Pair creation and annihilation rates. The solid line denotes the pair annihilation rate, the triangles denotes the pair creation rate, and $r$ is the radial coordinate.
}
\label{Pcre-ann}
\end{figure}

In Figure~\ref{Pcre-ann} we present the pair creation rate, which is calculated from the Monte Carlo simulation, and the pair annihilation rate, which is calculated from the physical quantities of electron-positron pairs. The pair annihilation rate decreases with increasing $r$ since the number densities of electrons $n_{\rm e^-}$ and positrons $n_{\rm e^+}$ decrease. The pair creation rate also decreases with $r$ because of the decrease of the number density of photons $n_{\rm \gamma}$ and the increase of its anisotropy. Since we assumed that electrons and positrons obey an isotropic Maxwell-Boltzmann distribution, the annihilation rate is simply obtained by equation~(\ref{3-04}). Let us compare the pair creation rate with the pair annihilation one. Inside the photosphere ($r<r_{\rm ph}$), the pair creation rate is almost equal to the annihilation one, indicating that pair creation balances pair annihilation. On the other hand, outside the photosphere ($r>r_{\rm ph}$) pair creation does not balance pair annihilation anymore. The pair creation rate is less than a thousandth of the pair annihilation rate and is negligible. 
The main reason for this is that the angular distribution of photons becomes anisotropic in this regime. 

\subsubsection{Radiative Force}

\begin{figure}[htp]
\centering
\epsscale{0.62}
\plotone{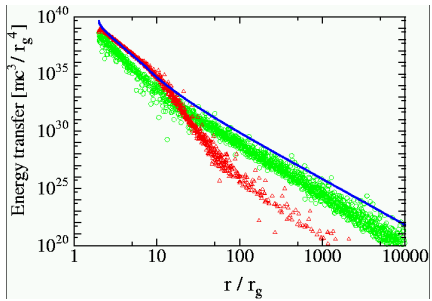}
\caption{
Energy transfer rates between photons and electron-positron pairs. This graph represents the energy transfer rates for three processes in the comoving frame of the pair plasma. The solid line denotes the energy loss of pairs due to pair annihilation, the open triangles denote the energy gain of pairs due to pair creation, and the open circles denote the energy gain of pairs through Compton scattering. 
}
\label{energy-transfer}
\end{figure}

Figure \ref{energy-transfer} shows the energy transfer rates between photons and pairs per unit comoving volume. The energy transfers through Compton scattering, pair annihilation, and pair creation are shown separately. The energy transfers for these three processes decrease with $r$ because the number densities of electrons, positrons, and photons decrease. The energy loss rate of pairs by pair annihilation is almost in proportion to the pair annihilation rate, and energy gain of pairs by pair creation is roughly proportional to the pair creation rate. In the optically thick regime, the energy loss and gain almost balance. The energy gain of pairs by Compton scattering is smaller than that by pair processes in the optically thick regime. In detail, the photons slightly give energy to pairs by these three processes ($F^0 >0$), but the radiative force is not as effective as the gas pressure force of the pairs themselves. The pairs and photons are bulk-accelerated by their own internal energy, and their luminosities do not vary much. In the optically thin regime, the energy gain by pair creation becomes smallest because of the steep decrease of the pair creation rate. 
\begin{figure}[htp]
\centering
\epsscale{0.62}
\plotone{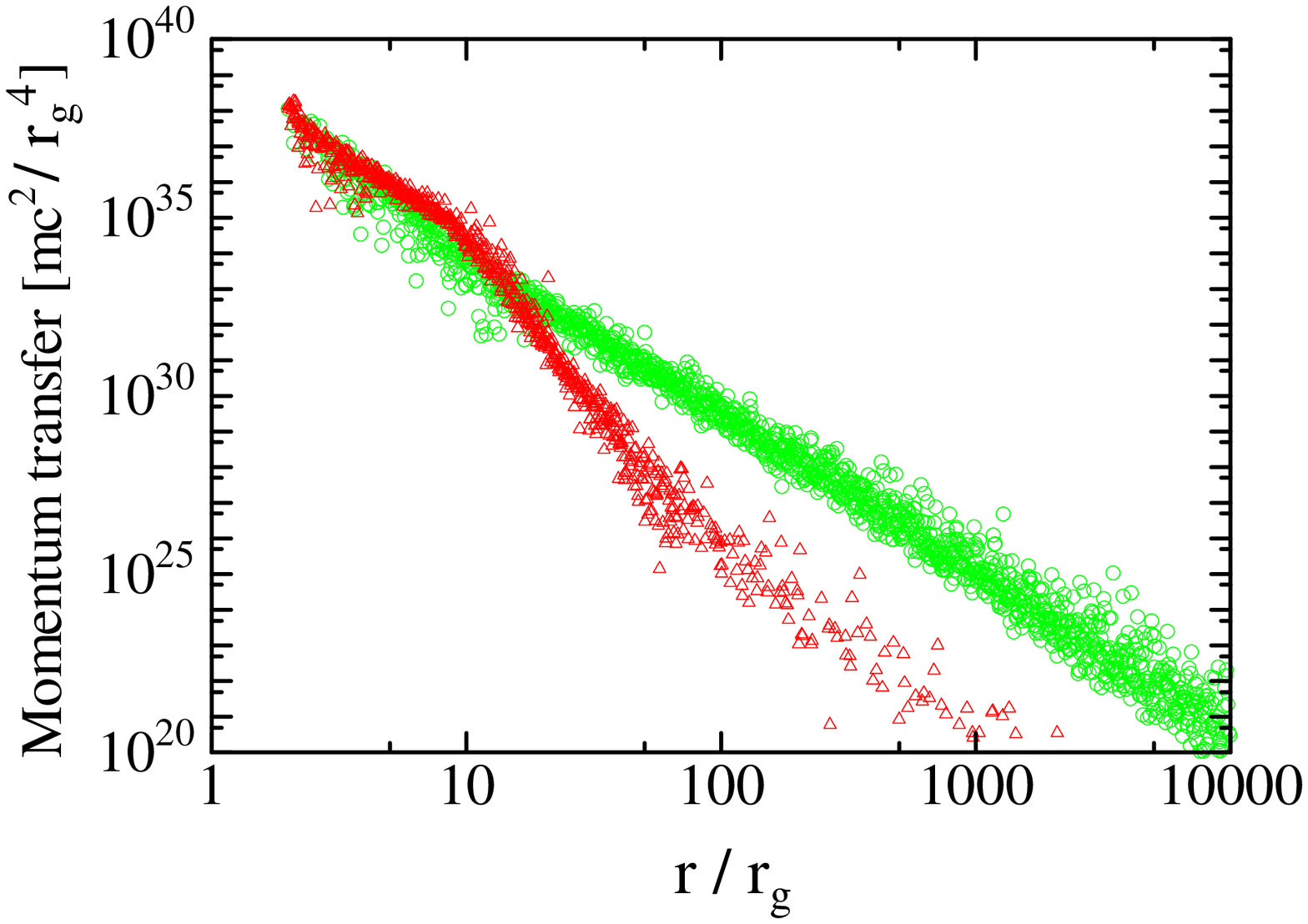}
\caption{
Momentum transfer rates from photons to electron-positron pairs in the comoving frame of pair plasma. The symbols are the same as in Figure \ref{energy-transfer}. The open triangles denote the momentum gain of pairs for pair creation, and the circles denote that for Compton scattering. There is no momentum transfer for pair annihilation. 
}
\label{momentum-transfer}
\end{figure}

Figure \ref{momentum-transfer} represents momentum transfers through pair creation and Compton scattering. The behavior is almost the same as the energy transfers shown in Figure \ref{energy-transfer}. Note that there is no momentum transfer for pair annihilation in the comoving frame of a fluid element because electrons and positrons are isotropic in the comoving frame. In both the optically thick and optically thin regimes, electron-positron pairs gain momentum from photons by pair creation and by Compton scattering (the same as the energy transfer). In the optically thick regime, the momentum gain by pair creation and that by Compton scattering are comparable to each other, but in the optically thin regime, the momentum gain by pair creation decreases faster because of a steep decrease in the pair creation rate. Therefore, the momentum gain by Compton scattering dominates and affects the pair plasma dynamics. As seen in Figure \ref{momentum-transfer}, photons give momentum to pairs and accelerate the pair plasma further, which indicates that radiation does not act as the drag force but does as the accelerating one.

\section{NUMERICAL RESULTS FOR VARIOUS BOUNDARY VALUES}
\label{NUMERICAL RESULTS FOR VARIOUS BOUNDARY VALUES}

We have also computed for various boundary values, and the results are tabulated in Table \ref{table2}. As is seen, the terminal Lorentz factor proves to be $5 \sim 20$ and the terminal kinetic luminosity of pairs accounts for about $40$\%-$60$\% of the total luminosity for $1 < \theta_0 < 4$ and $0.3 < \dot{E}/L_{\rm Edd} < 30$. Here we examine the dependence of the physical quantities at the photosphere on $\tau_0$ and show that they agree well with the analytic predictions ($r_{\rm ph} \propto {\tau_0}^{1/3} r_0$, $\theta_{\rm ph} \propto {\tau_0}^{-1/3} \theta_0$, and $\Gamma_{\rm ph} \propto {\tau_0}^{1/3}$; see \cite{Iwamoto02}). Then we describe the properties of the asymptotic values of the bulk Lorentz factor and the kinetic luminosity of the pair outflows. 

\begin{figure}[h]
\centering
\epsscale{0.65}
\plotone{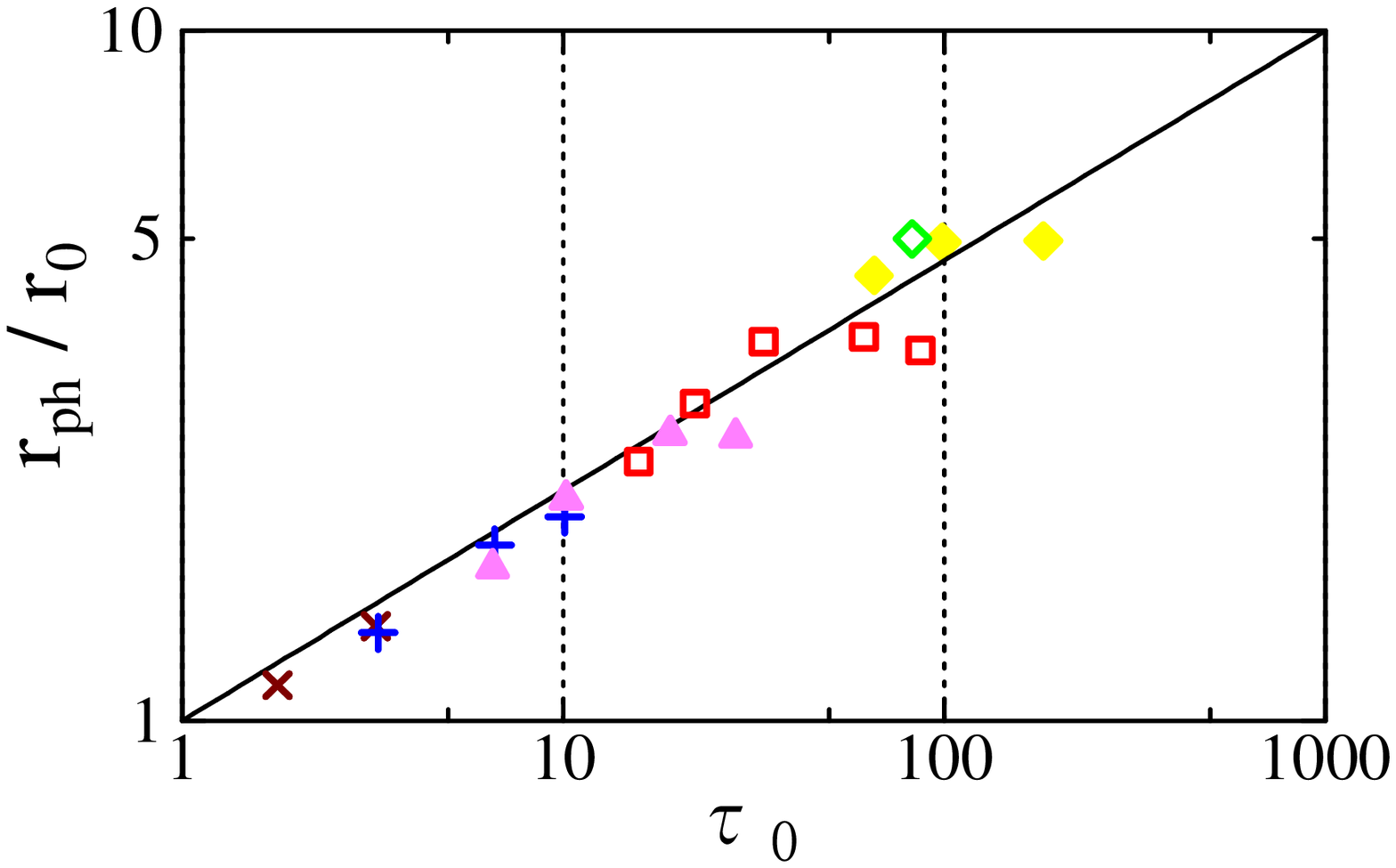}
\caption{
Relation between the photospheric radius $r_{\rm ph}$ and the initial optical thickness $\tau_0$. The solid line represents the analytic prediction $r_{\rm ph} / r_0 = {\tau_0}^{1/3}$. The crosses, plus signs, filled triangles, open squares, and filled diamonds refer to the cases $\dot{E}/ L_{\rm Edd}=$ $0.3$, $1$, $3$, $10$, and $30$, respectively. The open diamond refers to $\dot{E}/L_{\rm Edd} = 24.9$.
}
\label{photospheric-radius2}
\end{figure}

\begin{figure}[hbtp]
\centering
\epsscale{0.65}
\plotone{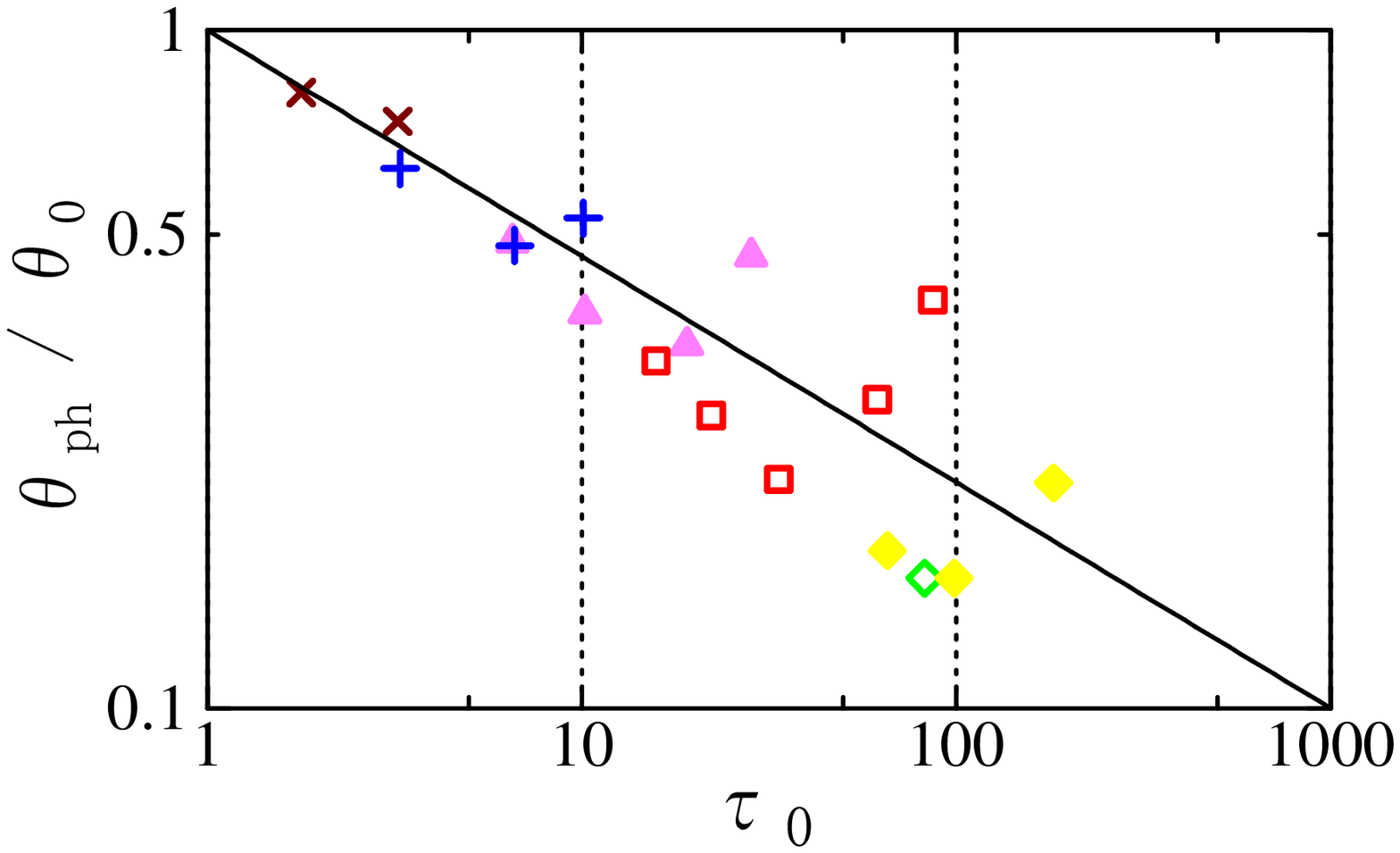}
\caption{
Relation between the photospheric temperature $\theta_{\rm ph}$ and the initial optical thickness $\tau_0$. The solid line represents the analytic prediction $\theta_{\rm ph}/ \theta_0 = {\tau_0}^{-1/3}$. The configuration of the symbols is the same as in Fig. \ref{photospheric-radius2}. 
}
\label{photospheric-temperature2}
\end{figure}

\begin{figure}[hbtp]
\centering
\epsscale{0.65}
\plotone{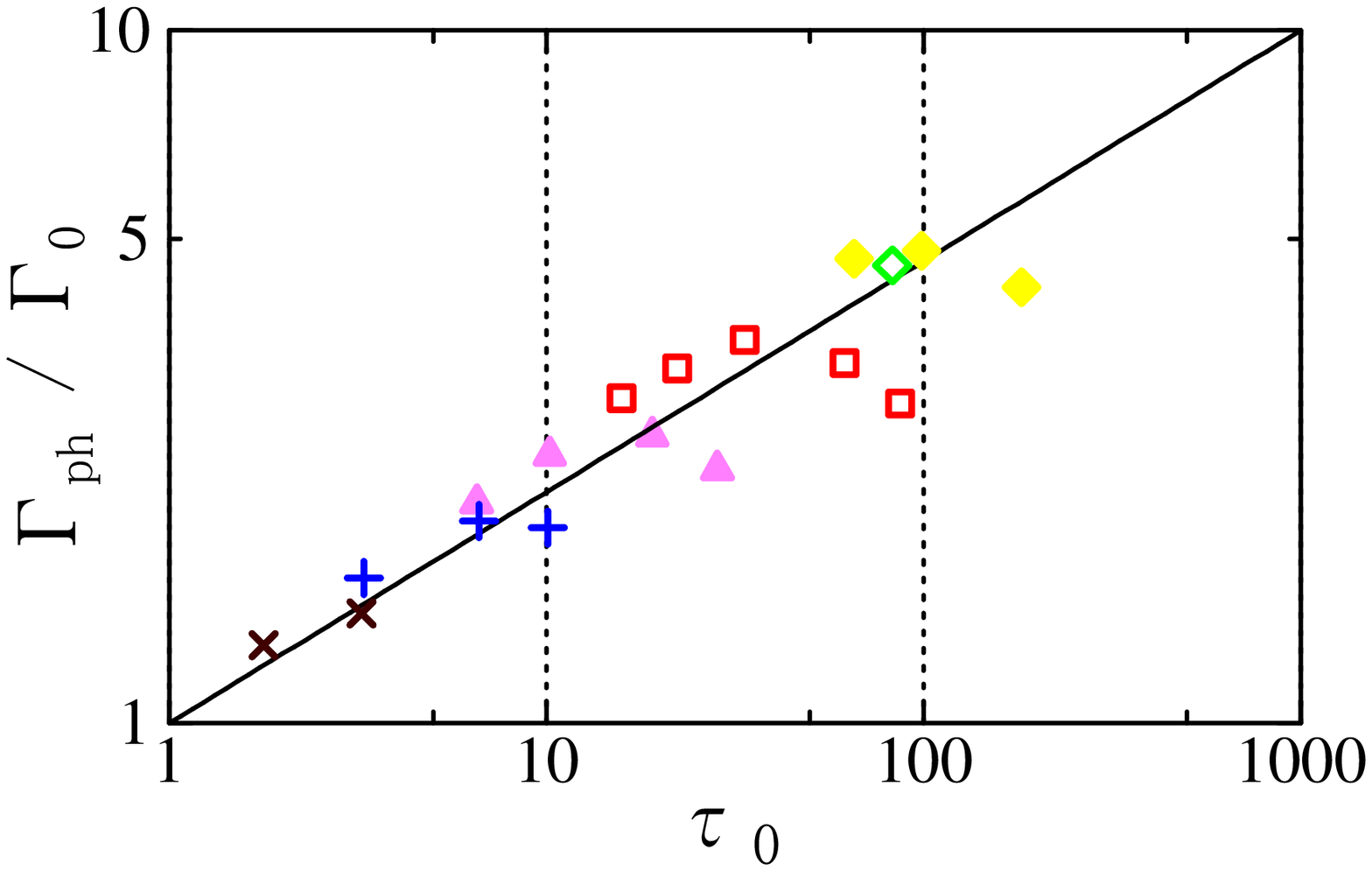}
\caption{
Relation between the photospheric Lorentz factor $\Gamma_{\rm ph}$ and the initial optical thickness $\tau_0$. The solid line represents the analytic prediction $\Gamma_{\rm ph} / \Gamma_0 = {\tau_0}^{1/3}$. The configuration of the symbols is the same as in Fig. \ref{photospheric-radius2}. 
}
\label{photospheric-velocity2}
\end{figure}

Figures \ref{photospheric-radius2}, \ref{photospheric-temperature2}, and \ref{photospheric-velocity2} show $r_{\rm ph}$, $\theta_{\rm ph}$, and $\Gamma_{\rm ph}$ as functions of $\tau_0$, respectively. We see that the numerical results agree well with the analytic predictions. As is shown in Figure \ref{photospheric-radius2}, $r_{\rm ph}$ is quite tightly correlated with $\tau_0$. In a few cases that correspond to the ``nonrelativistic freezeout,'' the numerical result deviates from the analytic prediction because a modest disappearance of pair plasma near the photosphere occurs at subrelativistic temperatures. Figure \ref{photospheric-temperature2} shows $\theta_{\rm ph}/\theta_0$ as a function of $\tau_0$. Numerical solutions roughly follow the analytic prediction, but they seem to give a slightly lower temperature than the analytic prediction for larger $\tau_0$. 
When $\tau_0$ is further increased for constant $\dot{E}$, $\theta_0$ and $\theta_{\rm ph}$ decreases, leading to the nonrelativistic freezeout. A few symbols that deviate from the line upward correspond to this case, where a lower temperature causes excess annihilation of pairs inside the photosphere. This leads to a smaller $r_{\rm ph}$ and larger $\theta_{\rm ph}/ \theta_0$ because the spatial interval for thermal expansion becomes shorter. Figure \ref{photospheric-velocity2} indicates that $\Gamma_{\rm ph}$ is also tightly correlated with $\tau_0$, and numerical results can be interpreted in the same way as in Figures \ref{photospheric-radius2} and \ref{photospheric-temperature2}. These features of the photosphere confirm the predictions of the thermal expansion inside the photosphere of Wien fireball even when we take account of detailed processes of radiation transfer such as the Klein-Nishina cross section and effects of anisotropic photon distribution. 

\vspace{0.6cm}

\begin{figure}[h]
\centering
\epsscale{0.65}
\plotone{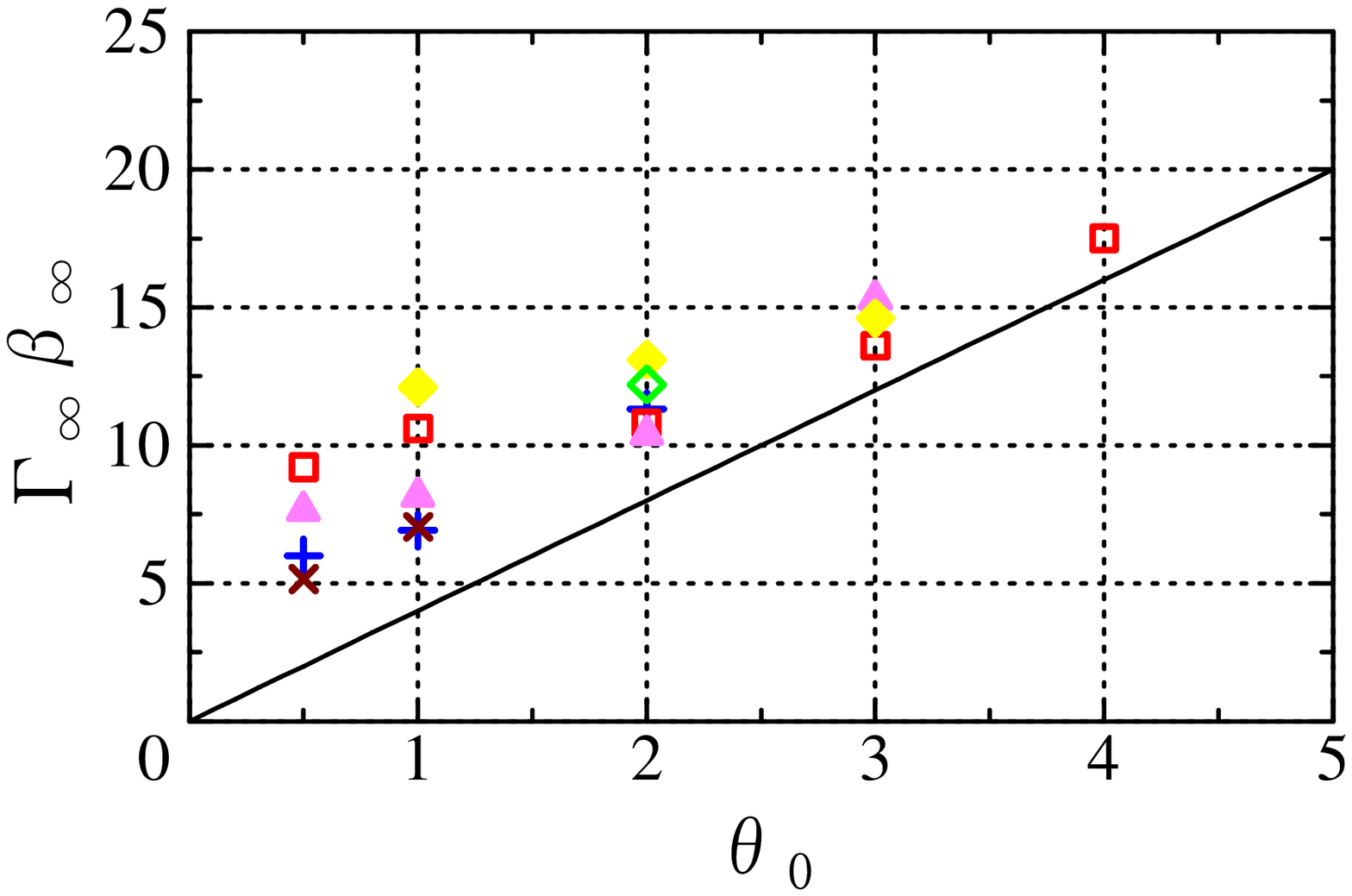}
\caption{
Terminal Lorentz factor $\Gamma_\infty \beta_\infty$. The abscissa is the temperature at the boundary $\theta_0$. The configuration of the symbols is the same as in Fig. \ref{photospheric-radius2}. 
}
\label{Ginf2}
\end{figure}

Next, we present the properties of the terminal outflow. In Figure \ref{Ginf2}, we present the terminal Lorentz factor. The terminal Lorentz factor mainly depends on the temperature at the inner boundary, since the electron-positron pairs are thermally accelerated both inside and outside the photosphere. The terminal Lorentz factor is analytically estimated as about $4\theta_0$ for the relativistic perfect fluid. It predicts that a higher boundary temperature results in a larger terminal Lorentz factor. In addition, beamed radiation from the photosphere further accelerates the pair plasma in the optically thin regime. Thus, the terminal Lorentz factor is larger than the simple analytic prediction of $4\theta_0$. For nonrelativistic freezeout ($\theta_{\rm ph} < 0.5$), this additional acceleration effect by radiation becomes large enough and $\Gamma_{\infty}$ does not decrease below $5$ even at a nonrelativistic temperature of $\theta_0=0.5$. 

\begin{figure}[htp]
\centering
\epsscale{0.65}
\plotone{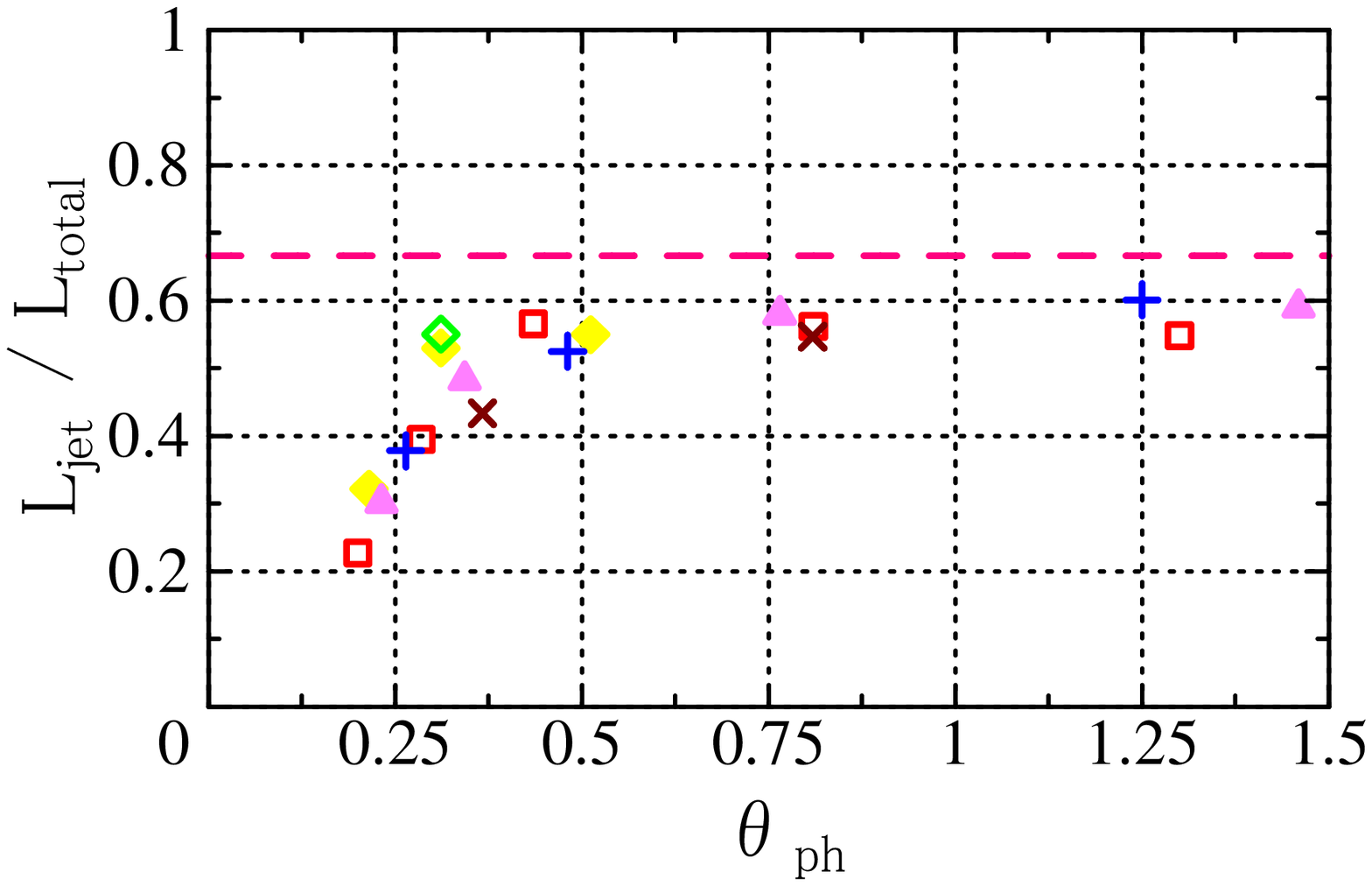}
\caption{
Ratio of the terminal kinetic luminosity of pairs to the total luminosity. The abscissa is the photospheric temperature. The configuration of the symbols is the same as in Fig. \ref{photospheric-radius2}, and the dashed line denotes the canonical value of $2/3$.
}
\label{Efficiency2}
\end{figure}

In Figure \ref{Efficiency2} we present the terminal kinetic luminosity divided by the total luminosity as a function of $\theta_{\rm ph}$. This ratio indicates the efficiency of energy conversion into the production of matter outflow. An efficiency of $2/3$ ({\it dotted line}) corresponds to the energy fraction of electron-positron pairs in a Wien equilibrium state at a relativistic temperature. For relativistic freezeout, the efficiency turns out to be about $50$\%-$60$\%, which is a little smaller than the canonical value of $2/3$. This is due to the energy loss for pair annihilation processes. For nonrelativistic freezeout, the efficiency becomes small because of the further decrease in pairs. The remaining part of the luminosity is accounted for by the MeV peaked radiation.

\hspace{0.4cm}


Summarizing these results, the attainable Lorentz factor of electron-positron pair plasma is more than $10$ for the inner boundary condition of $\theta_0>2$, and the kinetic luminosity is comparable to the  total luminosity, provided that Wien equilibrium states of pure electron-positron pairs are prepared at the inner boundary with a temperature of a few times the electron mass and an optical thickness to the scattering of more than $5$. When the photospheric temperature becomes nonrelativistic because of low boundary temperature and/or large optical thickness at the boundary, the terminal kinetic luminosity becomes low compared with the total luminosity, although the terminal Lorentz factor is still relativistic. The differences between the present results based on the detailed Monte Carlo simulation and the previous results based on simplified treatments of radiative transfer are surprisingly small. Generally, they are less than 10\%, and at most about $20$\%-$30$\%.


\section{CONCLUSION AND DISCUSSION}
\label{DISCUSSION}

We have considered relativistic outflow of electron positron pairs generated by the Wien fireball and obtained consistent numerical solutions of the dynamics of pairs and radiative transfer. We have utilized Monte Carlo simulation for a faithful treatment of radiative transfer including Compton scattering obeying the Klein-Nishina cross section, pair creation and annihilation, and we have obtained consistent solutions by several iterations. Compared with the results in our previous paper \citep{Iwamoto02}, in which simplified radiative transfer was adopted, the differences in numerical results are very small, generally less than 10\%, and at most about $20$\%-$30$\%. Thus, we have confirmed basic features of the Wien fireball model and the analytic predictions described in our previous paper. 
In addition, we have obtained a detailed energy and angular spectrum of photons, which enables us to calculate accurate radiative force. Basically, our present results indicate that our previous method overestimate the radiative force by about 10\%; thus, the correct terminal values of the Lorentz factor and kinetic luminosity become smaller while that of the number fraction of pairs becomes larger, compared with the results of the simplified treatments.

We briefly reiterate the basic features of the Wien fireball model. This model successfully achieves relativistic bulk acceleration and high kinetic power. It can avoid the problems with pair annihilation and radiation drag that have long been an annoyance with pair models. At the same time, it predicts a strong MeV peaked emission with a power similar to the kinetic power. 
These features are mainly due to the assumptions that a Wien fireball of electron positron pairs at relativistic temperature is produced and that the effects of external soft photons are neglected. Although whether such situations are realized or not is a critical issue for the present model, these issues have not been quantitatively discussed (see the discussion in \citet{Iwamoto02}). Our view is that instantaneous heating produces a runaway production of relativistic pairs due to slow annihilation and resultant dynamical expansion. The radiation field in the fireball is dominated by the internally produced gamma rays rather than external softer photons. For reference, the gamma-ray compactness of the fireball $l_{\rm \gamma} \sim {m_{\rm p} \over m_{\rm e}} {L \over L_{\rm Edd}} {r_g \over R}$ is very large, nominally about $10^4$-$10^5$, and about $10^2$-$10^3$ even when the collimation is taken into account (see below), as supposed. As a result, photons and pairs will quickly settle in a Wien equilibrium. When the fireball has a conical geometry in reality, the external soft photons will be reflected on the side of the fireball. The shielding process discussed by \citet{Illarionov96} can help to realize our situation since gamma rays from the fireball and X-rays from the outer disk make a pair atmosphere on the surface of the fireball and the atmosphere shields the penetration of soft photons into the fireball.

Since we have assumed spherical symmetry, we cannot solve the collimation problem, which should be investigated in the future with two- or three-dimensional treatments. Within spherical symmetry, the total luminosity of the successful Wien fireball must be higher than the Eddington luminosity. However, the required total luminosity becomes smaller if the outflow is collimated within a small opening angle, as is actually observed. Even when the jet is collimated to the solid angle $\Omega_{\rm j}$, the dynamics will not be much different from the spherical one and the total luminosity of jet becomes smaller by a factor $ \Omega_{\rm j}/4\pi$. Since $\Omega_{\rm j}$ is typically as small as $10^{-2}$, the required total luminosity can be smaller than the Eddington luminosity. 
As was discussed in our previous paper, there remain several unsolved issues with the Wien fireball model. These include the realization of the initial states in the hot accretion disk and its hot corona, or other ways to produce copious electron-positron pairs, the problem of baryon contamination, and collimation into the jet. These issues will be investigated in the future, along with observational confrontation of the Wien fireball model. 

\paragraph{Acknowledgements}
This work is supported in part by Grants-in-Aid for Scientific Research from the Ministry of Education and Science (F. T.; 13440061 and 14079205).

\begin{table}[htb]

\begin{center}
\begin{tabular}{|rrr|rrr|rrr|}

\hline

$\dot{E} / L_{\rm Edd}$ & $\theta_0$ & $\tau_0$ & $r_{\rm ph}$ & 
$\Gamma_{\rm ph} \beta_{\rm ph}$ & $\theta_{\rm ph}$ & 
$\Gamma_\infty \beta_\infty$ & $L_{\rm jet}$ & $L_{\rm jet} / \dot{E}$ \\

\hline
\hline

0.3 & 0.5 & 3.23 & 2.74 & 1.45 & 0.367 & 5.15 & 0.130 & 0.433 \\
0.3 & 1   & 1.78 & 2.25 & 1.23 & 0.809 & 7.05 & 0.164 & 0.547 \\

\hline

1.0 & 0.5 & 10.1 & 3.95 & 2.12 & 0.264 & 6.00 & 0.378 & 0.378 \\
1.0 & 1   & 6.63 & 3.59 & 2.18 & 0.481 & 6.91 & 0.525 & 0.525 \\
1.0 & 2   & 3.27 & 2.68 & 1.71 & 1.25 & 11.3 & 0.601 & 0.601 \\

\hline

3.0 & 0.5 & 28.4 & 5.15 & 2.66 & 0.232 & 7.63 & 0.904 & 0.301 \\
3.0 &   1 & 19.1 & 5.20 & 3.02 & 0.343 & 8.15 & 1.45 & 0.483 \\
3.0 &   2 & 10.2 & 4.19 & 2.81 & 0.765 & 10.4 & 1.74 & 0.580 \\
3.0 &   3 & 6.53 & 3.33 & 2.34 & 1.46 & 15.3 & 1.77 & 0.590 \\

\hline

10 & 0.5 & 86.8 & 6.88 & 3.40 & 0.200 & 9.20 & 2.27 & 0.227 \\
10 &   1 & 61.7 & 7.19 & 3.93 & 0.285 & 10.6 & 3.95 & 0.395 \\
10 &   2 & 33.6 & 7.08 & 4.26 & 0.435 & 10.8 & 5.66 & 0.566 \\
10 &   3 & 22.2 & 5.76 & 3.85 & 0.810 & 13.6 & 5.62 & 0.562 \\
10 &   4 & 15.8 & 4.75 & 3.46 & 1.30  & 17.5 & 5.48 & 0.548 \\

\hline

24.9 &   2 & 85.4 & 9.71 & 5.46 & 0.312 & 12.2 & 13.0 & 0.522 \\

\hline

30 &   1 & 182 & 9.91 & 5.11 & 0.215 & 12.1 & 9.65 & 0.322 \\
30 &   2 & 98.8 & 9.87 & 5.80 & 0.311 & 13.1 & 15.9 & 0.530 \\
30 &   3 & 65.5 & 8.83 & 5.64 & 0.512 & 14.6 & 16.5 & 0.550 \\

\hline

\end{tabular}

\caption{
Numerical results for various boundary conditions. First, second and third columns refer to the total luminosity, boundary temperature and boundary optical thickness. The fourth through sixth columns show the numerical results for the quantities at the photosphere, while the seventh through ninth columns show the numerical results for those at infinity.
\label{table2}
}

\end{center}

\end{table}

\end{document}